\documentclass[12pt,a4paper]{article}
\usepackage{amssymb}

\usepackage{amsmath}

\newcounter{resultnum}[section]

\newcounter{conclusionnum}[section]

\newcounter{conditionnum}[section]

\newcounter{conjecturenum}[section]

\newcounter{examplenum}[section]

\newcounter{exercisenum}[section]

\newcounter{lemmanum}[section]

\newcounter{notationnum}[section]

\newcounter{theoremnum}[section]

\newcounter{definitionnum}[section]

\newcounter{corollarynum}[section]

\newcounter{remarknum}[section]

\newcounter{propositionnum}[section]

\newcounter{acknowledgementnum}[section]

\newcounter{algorithmnum}[section]

\newcounter{axiomnum}[section]

\newcounter{casenum}[section]

\newcounter{claimnum}[section]

\newcounter{summarynum}[section]

\newcounter{problemnum}[section]

\begin{document}

\title{ Ricci Flows and Solitonic pp--Waves }

\author{ Sergiu I. Vacaru\thanks{%
Permanent:\
108-1490 Eginton Ave., Toronto, M6E 2G5, Canada
\newline  e-mails:
sergiu\_vacaru@yahoo.com, svacaru@brocku.ca  and svacaru@fields.utoronto.ca
 } \\
-- \\
\textsl{\ Department of Mathematics, Brock University,}\\
\textsl{St. Catharines, Ontario, Canada L2S\ 3A1} \\
and  \\
\textsl{The Fields Institute for Research in Mathematical
Science,}\\
\textsl{222 College Street, Second Floor,}\\
\textsl{ Toronto, Ontario, Canada M5T 3J1}\\
}

\date{September 20, 2006}
\maketitle

\begin{abstract}
We find exact solutions describing Ricci flows of four dimensional
pp--waves nonlinearly deformed by two/three dimensional solitons.
Such solutions are parametrized by five dimensional metrics with
gene\-ric off--diagonal terms and connections with nontrivial
torsion which can be related, for instance, to antisymmetric
tensor sources in string gravity. There are defined nontrivial
limits to four dimensional configurations and the Einstein gravity.

 \vskip2pt
Keywords:\  Ricci flows; gravitational solitons; pp--waves.

 \vskip2pt
 PACS numbers:\ 04.20.Jb, 04.30.Nk, 04.50.+h, 04.90.+e, 02.30.Jk

\end{abstract}

\section{Introduction}

Solutions with pp--waves \cite{ppwave1} and their generalizations are of
special interest in modern gravity and string theory in curved backgrounds %
\cite{strppwave}. In this paper, we consider three classes of five
dimensional (5D) metrics describing Ricci flows of solitonic
pp--waves (see Refs. \cite{gravsolit} and \cite{ricciflows} for
reviews of results  on gravitational solitons and, respectively,
Ricci flows). For non--solitonic configurations, a subclass of
such solutions has limits to the trivial embedding of the 4D
pp--wave metrics into 5D spacetimes.

There were elaborated various methods of constructing exact
solutions in gravity, see a summary of main approaches in Refs.
\cite{kramer} and \cite{bic}. The bulk of such solutions are with
a prescribed symmetry (spherical, cylindrical, torus, Killing or a
Lie algebra ones ...) and generated by  diagonalizable (by
coordinated transforms) ansatz for metrics. Such ansatz reduce the
Einstein equations to certain classes of  nonlinear ordinary
differential equations and/or algebraic equations which can be
solved exactly.   Then, an interesting discovery was the existence
of an infinite--dimensional group of transformations of metrics
mapping a solution of the Einstein equations to another solution
(Geroch group \cite{geroch}).

In our works \cite{anhm}, we developed a general geometric method of constructing
exact solutions by using anholonomic deformations of metrics  (they may
define, or not, any exact solutions) to certain classes of  generic
off--diagonal metrics exactly solving the gravitational field equations in
a model of gravity theory (we can consider the Einstein general relativity or
 various type of string, gauge,... both commutative and noncommutative generalizations).
  Such spacetimes, in general,  do not  possess any Killing/ group symmetry but
  can be characterized  by generalized
symmetries induced by 'anholonomy' relations  for preferred frames with
associated nonlinear connection structure. \footnote{%
The geometric constructions on spaces provided with  anholonomic structure
can be performed with resect to any  frames of reference and for any local
systems of coordinates. The term  'preferred' emphasizes that there are
certain classes of 'natural' frames  and coordinates when a symmetry of
fixed type can be distinguished in explicit form  and the field equations
can be exactly integrated.}  The exact off--diagonal solutions (for
instance, in 4D and/or 5D)\ constructed following the 'anholonomic frame
method' depend on classes of functions on 2--4 variables and can be
restricted to possess noncommutative symmetries, generalized Lie symmetries
like those for the Lie, or Clifford, algebroids. The papers \cite%
{anhm,details1,details2} contain details and references on recent
developments on constructing black ellipsoid and black torus solutions,
locally anisotropic wormholes, nonholonomically deformed Taub NUT spacetimes
and so on.

The aim of this paper is to show how we can apply the anholonomic frame method
in order to generate 5D exact solutions of the Ricci flows equations describing
nonlinear, for instance, solitonic deformations of 4D pp--waves\footnote{%
in a similar form we can consider generalizations to Kaigorodov spacetimes %
\cite{kaigorodov} and certain higher dimension solutions \cite{strppwave}}
as flows in 5D gravity. We shall define the conditions when such solutions
possess nontrivial limits to the Einstein gravity or their torsion can be
related to the antisymmetric 'H--field' in string gravity.

The paper is arranged as follows. After outlining some results on Ricci
flows for generic off--diagonal metrics in section 2, we consider 4D
pp--wave metrics and analyze their flows on a time like coordinate $v=z-t$
in 5D spaces with possible interactions with gravitational 2D solitons in
section 3. Then, in section 4, we construct exact solutions for Ricci flows
depending 'anisotropically' on coordinate $p=z+t$ when the metrics are
certain off--diagonal deformations from the a pp--wave background to
configurations with nonlinear interactions between pp--wave components and
3D solitonic waves. Section 5 is devoted to solutions describing flows
of pp--waves depending on the 5th coordinate when the torsion is related
completely antisymmetric tensor fields in sting gravity. Finally, in section
6, we conclude and discuss the obtained results.

\section{Off--Diagonal Ricci Flows}

We consider a linear quadratic element%
\begin{equation}
ds^{2}=g_{\alpha \beta }du^{\alpha }du^{\beta
}=g_{i}(x^{k})(e^{i})^{2}+g_{a}(x^{k},p)(e^{a})^{2}  \label{qel}
\end{equation}%
with the coefficients stated with respect to the co--frame%
\begin{equation}
e^{\alpha }=\left( e^{i}=dx^{i},e^{a}=dp^{a}+N_{i}^{a}(x^{k},p)dx^{i}\right)
\label{ddif}
\end{equation}%
being dual to the frame (funfbein)
\begin{equation}
e_{\alpha }=\left( e_{i}=\frac{\partial }{\partial x^{i}}-N_{i}^{a}\frac{%
\partial }{\partial p^{a}},e_{a}=\frac{\partial }{\partial p^{a}}\right) .
\label{dder}
\end{equation}%
In our formulas, the Einstein's summation rule on indices is adopted, the
indices of type $i,k,...$ run values $1,2,3;$ the indices of type $a,b,...$
run values $4,5$ and the local coordinates split in the form $u^{\alpha
}=(x^{i},p^{a}),$ where $p^{4}=p$ will be written in brief. The metric and
funfbein coefficients are parametrized by functions of necessary smooth
class,%
\begin{eqnarray}
g_{1} &=&\pm 1,g_{2,3}=q_{2,3}(x^{2},x^{3})\underline{g}%
_{2,3}(x^{2},x^{3}),g_{4,5}=q_{4,5}(x^{i},p)\underline{g}_{4,5}(x^{i}),
\notag \\
N_{i}^{4} &=&w_{i}(x^{i},p),N_{i}^{5}=n_{i}(x^{i},p).  \label{data1}
\end{eqnarray}%
We consider that the  coefficients $\underline{g}_{\alpha }$ are
given by a metric  in a theory of gravity (it may be a solution of
gravitational field equations  or, for instance, a conformal
transform of such a solution) and the coefficients $q_{\alpha
},w_{i}$ and $n_{i}$ are supposed to define deformations to other
classes of solutions. For our purposes, we can state any (pseudo)
Euclidean signature.

In this paper, we shall work both with the so--called 'canonical distinguished
connection' $\Gamma _{\ \beta \gamma }^{\alpha }=(L_{\
jk}^{i},L_{bk}^{a},C_{jc}^{i},C_{bc}^{a})$ and the Levi--Civita connection $%
\underline{\Gamma }_{\ \beta \gamma }^{\alpha },$ see details in Refs. \cite%
{details1,details2}. The coefficients of these connections can be computed
with respect to the local bases (\ref{dder}) and (\ref{ddif}), when
\begin{equation*}
\underline{\Gamma }_{\ \beta \gamma }^{\alpha }=\left( L_{\
jk}^{i},L_{bk}^{a}-\frac{\partial N_{k}^{a}}{\partial p^{b}},C_{jc}^{i}+%
\frac{1}{2}g^{ik}\Omega _{jk}^{a}g_{ca},C_{bc}^{a}\right)
\end{equation*}%
for%
\begin{eqnarray*}
\Omega _{ij}^{a} &=&e_{i}N_{j}^{a}-e_{j}N_{i}^{a},\ L_{\ jk}^{i}=\frac{1}{2}%
g^{ir}\left[ e_{k}(g_{jr})+e_{j}\left( g_{kr}\right) -e_{r}\left(
g_{jk}\right) \right] , \\
L_{bk}^{a} &=&e_{b}\left( N_{k}^{a}\right) +\frac{1}{2}g^{ac}\left[
e_{k}(g_{bc})-g_{dc}e_{b}(N_{k}^{d})-g_{db}e_{c}(N_{k}^{d})\right] , \\
C_{jc}^{i} &=&\frac{1}{2}g^{ik}e_{k}(g_{jk}),\ C_{bc}^{a}=\frac{1}{2}g^{ad}%
\left[ e_{c}(g_{bd})+e_{b}(g_{dc})-e_{d}(g_{bc})\right] .
\end{eqnarray*}%
The connection $\Gamma _{\ \beta \gamma }^{\alpha }$ is compatible with the
metric (\ref{qel}), i.e. it defines a covariant derivative $D_{\gamma }$
satisfying the conditions $D_{\gamma }g_{\alpha \beta }=0.$ The torsion $%
T_{\ \beta \gamma }^{\alpha }=\Gamma _{\ \beta \gamma }^{\alpha }-\Gamma _{\
\gamma \beta }^{\alpha }$ vanishes partially, $T_{\ jk}^{i}=0$ and $T_{\
bc}^{a}=0$   but there are nontrivial components
\begin{equation}
T_{ij}^{a}=\Omega _{ij}^{a},T_{ja}^{i}=C_{ja}^{i},T_{bj}^{a}=e_{b}\left(
N_{j}^{a}\right) -L_{bj}^{a}.  \label{tors}
\end{equation}%
By definition, the Levi--Civita connection is torsionless and metric
compatible, i.e. $\underline{T}_{\ \beta \gamma }^{\alpha }=0$ and $%
\underline{D}_{\gamma }g_{\alpha \beta }=0.$ We note that in order to
investigate flow solutions related to string gravity and spaces characterized by
 nonholonomic structures it is convenient to
use connections with nontrivial torsion. The constraints resulting
in zero torsion can be considered after certain general solutions
have been constructed.

The normalized Ricci flows \cite{ricciflows}, with respect to the coordinate
base $\partial _{\underline{\alpha }}=\partial /\partial u^{\underline{%
\alpha }},$ are described by the equations
\begin{equation}
\frac{\partial }{\partial \tau }g_{\underline{\alpha }\underline{\beta }%
}=-2R_{\underline{\alpha }\underline{\beta }}+\frac{2r}{5}g_{\underline{%
\alpha }\underline{\beta }},  \label{feq}
\end{equation}%
where the normalizing factor $r=\int RdV/dV$ is introduced in order to
preserve the volume $V.$ \footnote{%
we underlined the indices with respect to the coordinate bases in
order to distinguish them from those defined with respect to the
'N--elongated' local bases (\ref{dder}) and (\ref{ddif})} In this
work, we consider that the parameter $\tau $ may define flows on
any time like or extra dimension (5th) coordinate $x^{3},$ or $p.$
We note that we use the Ricci
tensor $R_{\underline{\alpha }\underline{\beta }}$ and scalar curvature $%
R=g^{\underline{\alpha }\underline{\beta }}R_{\underline{\alpha }\underline{%
\beta }}$ computed for the connection $\Gamma $ but further will shall
constrain the coefficients $N_{j}^{a}$ in such a manner that
\begin{equation}
R_{\alpha \beta }=\underline{R}_{\alpha \beta }  \label{econd}
\end{equation}%
in order to define limits to subclasses of solutions related to the Einstein
spaces. \footnote{%
the tensor $\underline{R}_{\underline{\alpha }\underline{\beta }}$ being
computed for the connection $\underline{\Gamma }$}

With respect to the funfbeins (\ref{dder}) and (\ref{ddif}), when
\begin{equation*}
e_{\alpha }=e_{\alpha }^{\ \underline{\alpha }}\ \partial _{\underline{%
\alpha }}\mbox{\ and \ }e^{\alpha }=e_{\ \underline{\alpha }}^{\alpha }\ du^{%
\underline{\alpha }}
\end{equation*}%
for the frame transforms respectively parametrized in the form%
\begin{equation*}
e_{\alpha }^{\ \underline{\alpha }}=\left[
\begin{array}{cc}
e_{i}^{\ \underline{i}}=\delta _{i}^{\underline{i}} & e_{i}^{\ \underline{a}%
}=N_{i}^{b}\ \delta _{b}^{\underline{a}} \\
e_{a}^{\ \underline{i}}=0 & e_{a}^{\ \underline{a}}=\delta _{a}^{\underline{a%
}}%
\end{array}%
\right] \mbox{\ and \ }e_{\ \underline{\alpha }}^{\alpha }=\left[
\begin{array}{cc}
e_{\ \underline{i}}^{i}=\delta _{\underline{i}}^{i} & e_{\ \underline{i}%
}^{b}=-N_{k}^{b}\ \delta _{\underline{i}}^{k} \\
e_{\ \underline{a}}^{i}=0 & e_{\ \underline{a}}^{a}=\delta _{\underline{a}%
}^{a}%
\end{array}%
\right] ,
\end{equation*}%
where $\delta _{\underline{i}}^{i}$ is the Kronecher symbol, the Ricci flow
equations (\ref{feq}) can be written
\begin{eqnarray}
\frac{\partial }{\partial \tau }g_{ii} &=&-2R_{ii}+2\lambda g_{ii}-g_{cd}%
\frac{\partial }{\partial \tau }(N_{i}^{c}N_{i}^{d}),\mbox{ for }i=2,3;
\label{eq1} \\
\frac{\partial }{\partial \tau }g_{aa} &=&-2R_{aa}+2\lambda g_{aa};%
\mbox{
for }a=4,5;\   \label{eq2} \\
R_{\alpha \beta } &=&0\mbox{ for }\alpha \neq \beta ,\text{ }  \notag
\end{eqnarray}%
where $\lambda =r/5$ and $g_{ij}$ and $g_{ab}$ are defined by the ansatz (%
\ref{qel}). The nontrivial components of the Ricci tensor $R_{\alpha \beta }$
are computed (see details of a similar calculus in Refs. \cite%
{details1,details2})
\begin{eqnarray}
R_{2}^{2} &=&R_{3}^{3}=\frac{1}{2g_{2}g_{3}}\left[ \frac{g_{2}^{\bullet
}g_{3}^{\bullet }}{2g_{2}}+\frac{(g_{3}^{\bullet })^{2}}{2g_{3}}%
-g_{3}^{\bullet \bullet }+\frac{g_{2}^{^{\prime }}g_{3}^{^{\prime }}}{2g_{2}}%
+\frac{(g_{2}^{^{\prime }})^{2}}{2g_{3}}-g_{2}^{^{\prime \prime }}\right] ,
\label{rth} \\
R_{4}^{4} &=&R_{5}^{5}=\frac{1}{2g_{4}g_{5}}\left[ -g_{5}^{\ast \ast }+\frac{%
\left( g_{5}^{\ast }\right) ^{2}}{2g_{5}}+\frac{g_{4}^{\ast }g_{5}^{\ast }}{%
2g_{4}}\right] ,  \label{rtv}
\end{eqnarray}%
where the functions $w_{i}$ and $n_{i}$ satisfy respectively the equations%
\begin{eqnarray}
-2g_{5}R_{4i} &=&w_{i}\beta +\alpha _{i}=0,  \label{eq3} \\
-2\frac{g_{4}}{g_{5}}R_{5i} &=&n_{i}^{\ast \ast }+\gamma n_{i}^{\ast }=0,
\label{eq4}
\end{eqnarray}%
for
\begin{eqnarray}
\alpha _{i} &=&\partial _{i}g_{5}^{\ast }-g_{5}^{\ast }\partial _{i}\ln
\sqrt{\left| g_{4}g_{5}\right| },\ \beta =g_{5}^{\ast \ast }-g_{5}^{\ast
}\left( \ln \sqrt{\left| g_{4}g_{5}\right| }\right) ^{\ast },  \label{aux1}
\\
\gamma &=&3g_{5}^{\ast }/2g_{5}-g_{4}^{\ast }/g_{4},\mbox{ for }g_{4}^{\ast
}\neq 0,g_{5}^{\ast }\neq 0,  \notag
\end{eqnarray}%
defined by $g_{4}$ and $g_{5}$ as solutions of equations (\ref{eq2}). In the
above presented formulas, it was convenient to write the partial derivatives
in the form $a^{\bullet }=\partial a/\partial x^{2},a^{^{\prime }}=\partial
a/\partial x^{3}$ and $a^{\ast }=\partial a/\partial p.$

In Refs. \cite{details1,details2} (see, for instance, the formulas
(160) and (161) in the Appendix to Ref. \cite{details2}), we proved
that for an ansatz of type (\ref{qel}) stated by data
(\ref{data1}) the conditions (\ref{econd}), generating Einstein
spaces, transform into certain systems of partial differential
equations for the off--diagonal metric coefficients
\begin{eqnarray}
2g_{5}w_{k}^{\ast \ast }+w_{k}^{\ast }g_{5}^{\ast } &=&0,  \label{econd1} \\
n_{k}^{\ast }g_{5}^{\ast } &=&0,  \notag
\end{eqnarray}%
and
\begin{eqnarray}
w_{2}^{\prime }-w_{3}^{\bullet }+w_{3}w_{2}^{\ast }-w_{2}w_{3}^{\ast } &=&0,
\label{econd2} \\
n_{2}^{\prime }-n_{3}^{\bullet }+w_{3}n_{2}^{\ast }-w_{2}n_{3}^{\ast } &=&0,
\notag
\end{eqnarray}%
Such constraints are necessary also in order to generate  Ricci flows of
 the Einstein metrics.

In the next sections, we shall construct explicit examples of exact
solutions, of equations (\ref{eq1})--(\ref{eq4}) for metric anzats of type (%
\ref{qel}) defined by data (\ref{data1}), describing flows of 4D metrics
into 5D spacetimes.

\section{Ricci flows on coordinate $\protect\tau =v$}

Let us consider a 5D metric
\begin{equation}
ds^{2}=-dx^{2}-dy^{2}+\frac{1}{8f(x,y,p)}dv^{2}-2f(x,y,p)\delta p^{2}\pm
\left( dp^{5}\right) ^{2},  \label{tsol1}
\end{equation}%
where
\begin{equation*}
\delta p=dp-\frac{1}{4f(x,y,p)}dv.
\end{equation*}%
For $p=z+t$ and $v=t-z,$ the ansatz (\ref{tsol1}) defines trivial extensions
on the 5th coordinate $p^{5}$ of the 4D pp--wave solutions of the vacuum
Einstein equations if the function $f(x,y,p)$ satisfies the condition%
\begin{equation*}
f_{xx}+f_{yy}=0,
\end{equation*}%
were $f_{x}=\partial f/\partial x,$ see the original paper \cite{ppwave1}
and recent developments in string theory \cite{strppwave}. \footnote{%
one can be performed both space or time like 5th dimension extensions}
Usually, one considers pp--waves parametrized in the form
\begin{equation}
f=f_{1}(x,y)\ f_{2}(p),  \label{ppar}
\end{equation}%
when, for instance, $f_{1}(x,y)\ =x^{2}-y^{2}$ and $f_{2}(p)=\sin p,$ or $%
f_{1}(x,y)\ =xy\left( x^{2}+y^{2}\right) $ and $f_{2}(p)=\exp
[b^{2}-p^{2}]^{-2}$ when $\mid p\mid <b$ and $f=0$ for $\mid p\mid \geq b.$

The pp--wave metric (\ref{tsol1}) is a particular case of off--diagonal
ansatz (\ref{qel}) when, for instance, the data (\ref{data1}) are stated by
nonzero coefficients
\begin{equation}
q_{\alpha }=1,\underline{g}_{1,2}=-1,\underline{g}_{3}=-1/8f,\underline{g}%
_{4}=-2f,\underline{g}_{5}=\pm 1,\underline{w}_{3}=-\frac{1}{4f}
\label{data2}
\end{equation}%
for the local coordinates parametrized in the form $x^{i}=(x,y,v)$ and $%
p^{a}=(p,p^{5}).$ In this section, we shall deform the metric (\ref{tsol1})
by introducing some values $q_{\alpha }(x^{i},p)$ and $N_{i}^{a}(x^{i},p),$
see (\ref{data1}), generating new solutions of the Ricci flow equations (\ref%
{eq1}), (\ref{eq2}) and (\ref{eq3}), (\ref{eq4}). We shall search solutions
for 'deformations' to an ansatz of type (\ref{tsol1}) stated in the form
\begin{eqnarray}
ds^{2} &=&-dx^{2}-g_{2}(v)\left[ dy^{2}-dv^{2}\right] -g_{4}(x,y,p)\delta
p^{2}\pm g_{5}(p)\left( \delta p^{5}\right) ^{2},  \label{tsol2} \\
\delta p &=&dp+w_{2}(x,y)dy+w_{3}(p)dv,  \notag \\
\delta p^{5} &=&dp^{5}+n_{i}(x,y,v,p)dx^{i}.  \notag
\end{eqnarray}%
The coefficients from (\ref{tsol2}) are defined by multiplying the data (\ref%
{data2}) on respective 'polarization' functions when
\begin{eqnarray}
g_{2}(v) &=&\eta _{2}(y,v)\underline{g}_{2},g_{3}(v)=\eta _{3}(x,y,v,p)%
\underline{g}_{3}=g_{2}(v),  \label{data3} \\
g_{4}(x,y,p) &=&\eta _{4}(x,y,v,p)\underline{g}_{4},\ g_{5}(p)=\eta _{5}(p)%
\underline{g}_{5},\ w_{3}(p)=\eta _{3}^{[w]}(x,y,v,p)\underline{w}_{3},
\notag
\end{eqnarray}%
and by considering certain additional/ modified 'N--coefficients' $%
w_{2}(x,y) $ and $n_{i}(x,y,p).$

Computing the components $R_{ij}$ of the Ricci tensor (\ref{rth}) for the
connection $\Gamma _{\ \beta \gamma }^{\alpha }$ defined by ansatz (\ref%
{tsol2}), we write the Ricci flow equation (\ref{eq1}) in the form
\begin{equation}
\ \theta ^{\prime \prime }+\frac{1}{\theta }(\theta ^{\prime })^{2}-\theta \
\theta ^{\prime }+2\lambda \theta =0,  \label{bern}
\end{equation}%
where $\theta (v)=g_{2}(v)=g_{3}(v).$ This equation is investigated in Ref. %
\cite{kamke}. Stating $\mu (\theta )=d\theta (v)/dv,$ we obtain the Abel
equation
\begin{equation*}
\mu \ \frac{d\mu }{d\theta }+\frac{1}{\theta }\mu ^{2}-\theta \mu +2\lambda
\theta =0.
\end{equation*}%
A similar equation, without the term $+\frac{1}{\theta }\mu ^{2},$
\begin{equation*}
\mu \ \frac{d\mu }{d\theta }-\theta \mu +2\lambda \theta =0,
\end{equation*}%
was considered for various classes of exact off--diagonal solutions in
gravity, see discussions and references in Refs. \cite{details1,details2}. For $%
\lambda \rightarrow 0,$ we get a Bernoulli type equation. We conclude that
for the ansatz (\ref{tsol2}) the Ricci flow of components $g_{2}(v)=g_{3}(v)$
are governed by explicit solutions of the Abel or Bernoulli equations.

For a subclass of solutions when $g_{4,5}$ and $N_{i}^{a}$ in (\ref{tsol2})
do not depend on variable $x^{3}=v,$ the terms $\frac{\partial }{\partial
\tau }g_{aa}$ in (\ref{eq2}) vanish, i.e. $R_{4}^{4}=R_{5}^{5}=\lambda ,$
\begin{equation}
g_{5}^{\ast }\left[ -\frac{g_{5}^{\ast \ast }}{g_{5}^{\ast }}+\frac{%
g_{5}^{\ast }}{2g_{5}}+\frac{g_{4}^{\ast }}{2g_{4}}\right] =2g_{4}
g_{5}\lambda .  \label{eq2a}
\end{equation}%
The general procedure of constructing exact solutions of the systems of
equations (\ref{eq2a}), (\ref{eq3}) and (\ref{eq4}) is described in
Appendices to Refs. \cite{details1,details2}. Here we present the final
results of such computations describing 4D generic off--diagonal Einstein
metrics (by straightforward computations one can be verified that the given
sets of functions really solve the field equations): One has%
\begin{equation}
g_{4}=\zeta \eta _{4}^{[\lambda =0]}\underline{g}_{4}\mbox{\ and \ }%
g_{5}=\eta _{5}(p)\underline{g}_{5},  \label{data4a}
\end{equation}%
where%
\begin{eqnarray}
\zeta (x,y,p) &=&1+\frac{\lambda }{8}h_{0}^{2}(x,y)\int \sqrt{\mid \eta
_{5}\ \eta _{5}^{\ast }\mid }dp,  \label{data4c} \\
\eta _{4}^{[\lambda =0]}(x,y,p) &=&\frac{h_{0}^{2}(x,y)}{f_{1}(x,y)}\left[
f_{2}^{\ast }-\frac{f_{2}f_{2}^{\ast \ast }}{2\left( f_{2}^{\ast }\right)
^{2}}\right] ,\ \eta _{5}(p)=\left( f_{2}\right) ^{4}/\left( f_{2}^{\ast
}\right) ^{2}  \notag
\end{eqnarray}%
for an arbitrary function $h_{0}(x,y),$ a given $\underline{g}%
_{4}=-2f_{1}(x,y)f_{2}(p)$ defined by the 'non--deformed' pp--wave solution,
and for $\underline{g}_{5}=\pm 1,$ depending on the signature of the 5th
dimension. We can satisfy the conditions (\ref{econd1}) and (\ref{econd2})
by any functions
\begin{equation}
w_{2}=w_{2}(x,y)\mbox{\ and \ }w_{3}=\left[ f_{2}(p)\right] ^{-1}
\label{data4d}
\end{equation}%
and any
\begin{equation}
n_{2}=n_{2}(x,y,v)\mbox{\ and \ }n_{3}=n_{3}(x,y,v)  \label{data4e}
\end{equation}%
for which $n_{2}^{\prime }-n_{3}^{\bullet }=0$ and $n_{2,3}^{\ast }=0.$

In explicit form, such Ricci flow solutions are described by the generic
off--diagonal metric
\begin{eqnarray*}
ds^{2} &=&-dx^{2}-\theta (v)\left[ dy^{2}-dv^{2}\right]
-2h_{0}^{2}(x,y)f_{2}(p)\left[ f_{2}^{\ast }-\frac{f_{2}f_{2}^{\ast \ast }}{%
2\left( f_{2}^{\ast }\right) ^{2}}\right] \times \\
&&\left( 1+\frac{\lambda }{4}h_{0}^{2}(x,y)\int \left( f_{2}\right) ^{3}%
\sqrt{\mid \frac{f_{2}}{f_{2}^{\ast }}-\frac{\left( f_{2}\right) ^{2}}{%
2\left( f_{2}^{\ast }\right) ^{3}}\mid }dp\right) \delta p^{2} \\
&&\pm \left( f_{2}\right) ^{4}/\left( f_{2}^{\ast }\right) ^{2}\left( \delta
p^{5}\right) ^{2}, \\
\delta p &=&dp+w_{2}(x,y)dy+\left[ f_{2}(p)\right] ^{-1}dv, \\
\delta p^{5} &=&dp^{5}+n_{2}(x,y,v)dy+n_{3}(x,y,v)dv.
\end{eqnarray*}

By straightforward computations, we can verify that the coefficients $%
g_{2}(v)$ $=g_{3}(v),$ solving (\ref{bern}), and the coefficients (\ref%
{data4a})--(\ref{data4e}) for the ansatz (\ref{tsol2}) define a class of
exact solutions of equations (\ref{eq1}) and (\ref{eq2}) describing
off--diagonal Ricci flows of 4D Einstein metrics into certain 5D spacetimes.
The solutions depend on arbitrary integration functions $h_{0}(x,y),$ $%
w_{2}(x,y)$ and $n_{2,3}(x,y,v),$ for which $n_{2}^{\prime }=n_{3}^{\bullet
}.$ We emphasize that such functions can not be completely eliminated by
coordinate transforms:\ they describe certain classes of spacetimes which
can be further constrained to satisfy explicit symmetry conditions for
nonlinear gravitational interactions and flows. For instance, one can be
considered the case when $h_{0}(x,y)$ is a solution of a 2D solitonic
equation, for instance, of the sine--Gordon equation (SG),%
\begin{equation*}
\pm \frac{\partial ^{2}h_{0}}{\partial x^{2}}\mp \frac{\partial ^{2}h_{0}}{%
\partial y^{2}}=\sin (h_{0}).
\end{equation*}%
This way we generate a class of solutions describing nonlinear off--diagonal
interactions of gravitational pp--waves and 2D solitonic waves with
nontrivial Ricci flows of some components of metric satisfying the
corresponding Bernoulli, or Abel, type equation. If $g_{2}^{\prime
}(v_{0})=g_{3}^{\prime }(v_{0})=0$ in a point $v=v_{0},$ the flow solutions
are approximated by a set of exact off--diagonal solutions of the vacuum
Einstein equations with nontrivial cosmological constant defining solitonic
pp--waves. This class of metrics is generated by corresponding off--diagonal
deformations of the 4D pp--wave metric embedded into 5D spacetime, see
ansatz (\ref{tsol1}).

We can generalize our solutions by introducing certain dependencies on the
''flow'' coordinate $v$ in $w_{2,3}(x,y,v)$ $\ $and $n_{2,3}(x,y,v)$ not
subjected to the conditions (\ref{econd1}) and (\ref{econd2}), for instance,
by supposing that some such components are solutions of the 3D solitonic
equations. This way, one generates Ricci flows of pp--waves nonlinearly
interacting with 3D solitons inducing nontrivial torsion (\ref{tors}) which
under corresponding conditions can be related to a nontrivial torsion field
in string gravity, Einstein--Cartan gravity, or certain models of
metric--affine, Finsler and/or noncommutative generalizations of gravity,
see Refs. \cite{details1,details2} and the section \ref{flowsg} in this work.

\section{Ricci flows on coordinate $\protect\tau =p$}

We analyze an ansatz
\begin{eqnarray}
ds^{2} &=&-dx^{2}+g_{2}(x^{i})dy^{2}+g_{3}(x^{i})dv^{2}+g_{4}(x^{i},p)\left(
\delta p\right) ^{2}+g_{5}(x^{i},p)\left( \delta p^{5}\right) ^{2},  \notag
\\
\delta p &=&dp+w_{3}(x^{i},p)dv,\delta p^{5}=dp^{5}+n_{3}(x^{i},p)dv,
\label{tsol3}
\end{eqnarray}%
for $g_{4,5}=\eta _{4,5}(x,y,v,p)\underline{g}_{4,5},$ with $\underline{g}%
_{4}=-2f_{1}(x,y)f_{2}(p)$ and $\underline{g}_{4}=\pm 1,$ defined by a
solution of pp--wave equation (\ref{ppar}), when the local coordinates are
stated in the form $x^{i}=(x,y,v)$and $p^{a}=(p,p^{5}).$ We search for
solutions of the system of equations (\ref{eq1}), (\ref{eq2}) and (\ref{eq3}%
), (\ref{eq4}) beginning with solutions of $R_{a}^{a}=0,$ see (\ref{rtv}),
of type
\begin{equation}
\sqrt{\mid g_{4}\mid }=h_{0}\left( \sqrt{\mid g_{5}\mid }\right) ^{\ast
},h_{0}=const,  \label{ae1}
\end{equation}%
when the term $g_{cd}\frac{\partial }{\partial p}(N_{i}^{c}N_{i}^{d})$ in (%
\ref{eq1}) is constrained to be zero as the solution of the equation
\begin{equation}
g_{4}\frac{\partial }{\partial p}(w_{3})^{2}+g_{5}\frac{\partial }{\partial p%
}(n_{3})^{2}=0,  \label{ae2}
\end{equation}%
for $N_{i}^{4}=w_{i}$ and $N_{i}^{5}=n_{i}$ where$\frac{\partial }{\partial p%
}w_{3}=w_{3}^{\ast }(x,y,v,p).$ If the relation (\ref{ae1}) holds, one has $%
\alpha _{i}=\beta =0,$ see formulas (\ref{aux1}). This allows us to consider
any values $w_{i}$ in (\ref{eq3}). For instance, we can take $%
w_{3}(x,y,v,p)\neq 0$ and $w_{1}=w_{2}=0$ \footnote{%
for simplicity, we chose a 'minimal' extension of the set of such functions}%
.\ For the ansatz (\ref{tsol3}), parametrizing $\eta _{a}=\eta
_{a}^{[1]}(x,y,v)\eta _{a}^{[2]}(p),$ one obtains from (\ref{ae1}) that%
\begin{equation*}
\eta _{5}^{[1]}=\mid 2f_{1}\eta _{4}^{[1]}\mid
\end{equation*}%
and
\begin{equation*}
\sqrt{\mid 2f_{2}(p)\ \eta _{4}^{[2]}(p)\mid }=h_{0}\left( \sqrt{\mid \eta
_{5}^{[2]}(p)\mid }\right) ^{\ast }.
\end{equation*}%
This is compatible with the equation (\ref{eq2}) if $\sqrt{\mid 2\lambda
\mid }\ h_{0}=1,$ when%
\begin{equation*}
\eta _{5}^{[2]}(p)=\eta _{5[0]}e^{2\lambda p}=f_{2}(p)\ \eta _{4}^{[2]}(p),
\end{equation*}%
for $\eta _{5[0]}=const.$

A nontrivial solution of (\ref{eq4}) and (\ref{ae1}) can be written in the
form%
\begin{equation}
n_{1,2}=0,n_{3}=n_{3[1]}(x,y,v)+n_{3[2]}(x,y,v)\int \frac{e^{-\lambda p}dp}{%
\sqrt{\mid f_{2}(p)\mid }},  \label{aux01}
\end{equation}%
where $n_{3[1,2]}(x,y,v)$ are some integration functions. We can compute $%
w_{3}$ by integrating (\ref{ae2}) for defined $g_{4,5}$ and $n_{3},$%
\begin{equation}
w_{3}=\left| 2f_{1}(x,y)\left[ f_{2}(p)-\int f_{2}(p)dp\right] \right|
^{1/2}\ n_{3}(x,y,v,p).  \label{aux02}
\end{equation}

The final step in constructing this class of solutions is to solve the
equation (\ref{eq1}) which for the ansatz (\ref{tsol3}), see formulas (\ref%
{rtv}), and the condition (\ref{ae2}) transforms into
\begin{equation*}
\frac{g_{2}^{\bullet }g_{3}^{\bullet }}{2g_{2}}+\frac{(g_{3}^{\bullet })^{2}%
}{2g_{3}}-g_{3}^{\bullet \bullet }+\frac{g_{2}^{^{\prime }}g_{3}^{^{\prime }}%
}{2g_{2}}+\frac{(g_{2}^{^{\prime }})^{2}}{2g_{3}}-g_{2}^{^{\prime \prime
}}=2g_{2}g_{3}\lambda .
\end{equation*}%
The general solution of this equation can be written in the form%
\begin{equation*}
\varpi =g_{[0]}\exp \left[ a_{2}\ \widetilde{x}^{2}(y,v)+a_{3}\ \widetilde{x}%
^{3}(y,v)\right] ,
\end{equation*}%
where $g_{[0]}$ and $a_{2,3}$ are some constants and the functions $%
\widetilde{x}^{2,3}(y,v)$ define a coordinate transform $x^{2,3}\rightarrow
\widetilde{x}^{2,3}$ for which the 2D line element becomes conformally flat,%
\begin{equation*}
g_{2}(x^{i})dy^{2}+g_{3}(x^{i})dv^{2}\rightarrow \varpi (y,v)\left[ \left( d%
\widetilde{x}^{2}\right) ^{2}+\left( d\widetilde{x}^{3}\right) ^{2}\right]
\end{equation*}%
with $\psi =\ln \mid \varpi \mid $ satisfying the equation
\begin{equation}
\psi ^{\bullet \bullet }-\psi ^{\prime \prime }=-2\lambda .  \label{aux03}
\end{equation}

Summarizing the coefficients for the ansatz (\ref{tsol3}), we get the metric
\begin{eqnarray}
ds^{2} &=&-dx^{2}-e^{\psi (y,v)}\left[ \left( d\widetilde{x}^{2}(x,y)\right)
^{2}-\left( d\widetilde{x}^{3}(x,y)\right) ^{2}\right]  \label{tsol3a} \\
&&-2f(x,y,p)\eta _{4}^{[1]}(x,y,v)\eta _{4}^{[2]}(p)\left( \delta p\right)
^{2}\pm \eta _{5}^{[0]}e^{2\lambda p}\left( \delta p^{5}\right) ^{2},  \notag
\\
\delta p &=&dp+w_{3}(x,y,v,p)dv,\delta p^{5}=dp^{5}+n_{3}(x,y,v,p)dv,  \notag
\end{eqnarray}%
depending on 'polarization' (integration) functions $\eta _{4}^{[1]}$ and $%
n_{3[1,2]}$ and constant $\eta _{5}^{[0]}$ with nontrivial N--coefficients $%
n_{3}$ and $w_{3}$ defined respectively by the equations (\ref{aux01}) and (%
\ref{aux02}) and $\psi (y,v)$ being a solution of (\ref{aux03}). We can
state in explicit form a nonlinear superposition of a pp--wave $f=f_{1}f_{2}$
and a 3D soliton when $\eta _{4}^{[1]}(x,y,v)$ is a solution, for instance,
of the (2+1)--dimensional SG equation,%
\begin{equation*}
-\left( \eta _{4}^{[1]}\right) ^{\bullet \bullet }+\left( \eta
_{4}^{[1]}\right) ^{\prime }+\frac{\partial ^{2}}{\left( \partial
x^{1}\right) ^{2}}\left( \eta _{4}^{[1]}\right) =\sin \left( \eta
_{4}^{[1]}\right) .
\end{equation*}

It should be noted that the metric (\ref{tsol3a}) describes solitonic
pp--wave flows for the Ricci tensor defined for the canonical distinguished
connection $\Gamma _{\ \beta \gamma }^{\alpha }$ possessing certain
nontrivial torsion coefficients. We can restrict the integral varieties (of
solutions) by considering only some  nontrivial coefficients $w_{3}(v)$ and $%
n_{3}(v)$ depending on variable $x^{3}=v$ with the integration functions
subjected to the conditions (\ref{econd1}) and (\ref{econd2}). In such
cases, we extract exact solutions for flows (on coordinate $p=z+t)$ of a 4D\
Einstein space propagating as a nontrivial embedding into 5D spacetimes.

\section{Flows on Extra Dimension Coordinate}

\label{flowsg}Other classes of exact solutions for Ricci flows can be
constructed by deforming the metric (\ref{tsol1}) as in previous section but
with reparametrized coordinates $x^{i}=(x,y,v)$ and $p^{a}=(p^{5},p),$ by
multiplying the coefficients $\underline{g}_{\alpha }$ on polarization
functions%
\begin{eqnarray*}
\eta _{1} &=&1,\eta _{2}=\eta _{2}(y,v),\eta _{3}=f_{2}(p)\eta
_{3}^{[1]}(y,v), \\
\eta _{4} &=&\left[ f_{2}(p)\right] ^{-1}\eta _{4}^{[1]}(x,y,v)\eta
_{4}^{[2]}(p^{5}),\eta _{5}=\eta _{5}(x,y,v,p^{5})
\end{eqnarray*}%
and then following similar geometric constructions when the dependence on
coordinate $p^{5}$ for a pp--wave $\ ^{ext}f(x,y,p^{5})$ is emphasized
instead of a usual function $f(x,y,p)$ for 4D solutions. The derived
formulas and equations are similar but with $p\longleftrightarrow p^{5}$
defining flows on the 5th coordinate $p^{5}.$ Such solutions are
parametrized by the ansatz
\begin{eqnarray}
ds^{2} &=&-dx^{2}-e^{\psi (y,v)}\left[ \left( d\widetilde{x}^{2}(x,y)\right)
^{2}-\left( d\widetilde{x}^{3}(x,y)\right) ^{2}\right]  \label{tsol4} \\
&&-2f(x,y,p^{5})\eta _{4}^{[1]}(x,y,v)\eta _{4}^{[2]}(p^{5})\left( \delta
p^{5}\right) ^{2}\pm \eta _{5}^{[0]}e^{2\lambda p^{5}}\left( \delta p\right)
^{2},  \notag \\
\delta p^{5} &=&dp^{5}+w_{3}(x,y,v,p^{5})dv,\ \delta
p=dp+n_{3}(x,y,v,p^{5})dv,  \notag
\end{eqnarray}%
with the coefficients defined by equations of type (\ref{aux01}), (\ref%
{aux02}) and (\ref{aux03}), and can be constrained, for certain $w_{3}(v)$
and $n_{3}(v)$ to define flows of the Einstein spaces for the Levi--Civita
connection $\underline{\Gamma }.$

For solutions with nontrivial torsion (\ref{tors}), we can relate the Ricci
tensor flows to a source defined by the strength (torsion)
\begin{equation*}
H_{\mu \nu \rho }=e_{\mu }B_{\nu \rho }+e_{\rho }B_{\mu \nu }+e_{\nu
}B_{\rho \mu }
\end{equation*}%
of an antisymmetric field $B_{\nu \rho },$ when%
\begin{equation}
R_{\mu \nu }=-\frac{1}{4}H_{\mu }^{\ \nu \rho }H_{\nu \lambda \rho }
\label{c01}
\end{equation}%
and
\begin{equation}
D_{\lambda }H^{\lambda \mu \nu }=0,  \label{c02}
\end{equation}%
see details on string gravity, for instance, in Refs. \cite{string}. \ Here,
we also note that the 3D string gravity is used for a modified Ricci flow
analysis of the Thurston geometrization conjecture \cite{geg} but our
approach is oriented to constructing exact solutions. The conditions (\ref%
{c01}) and (\ref{c02}) are satisfied by the ansatz
\begin{equation}
H_{\mu \nu \rho }=\widehat{Z}_{\mu \nu \rho }+\widehat{H}_{\mu \nu \rho
}=\lambda _{\lbrack H]}\sqrt{\mid g_{\alpha \beta }\mid }\varepsilon _{\nu
\lambda \rho }  \label{ansh}
\end{equation}%
where $\varepsilon _{\nu \lambda \rho }$ is completely antisymmetric and the
distorsion (from the Levi--Civita connection) $\widehat{Z}_{\mu \nu \rho }$
is defined by the torsion tensor (\ref{tors}). \footnote{%
We emphasize that our H--field ansatz is different from those already used
in string gravity when $\widehat{H}_{\mu \nu \rho }=\lambda _{\lbrack H]}%
\sqrt{\mid g_{\alpha \beta }\mid }\varepsilon _{\nu \lambda \rho }$ \cite%
{string}. \ In our approach, we define $H_{\mu \nu \rho }$ and $\widehat{Z}%
_{\mu \nu \rho }$ from respective ansatzes for the H--field and
nonholonomically deformed metric, compute the torsion tensor for the
canonical distinguished connection $\Gamma $ and after that it is possible
to find the 'deformed' H--field, $\widehat{H}_{\mu \nu \rho }=\lambda
_{\lbrack H]}\sqrt{\mid g_{\alpha \beta }\mid }\varepsilon _{\nu \lambda
\rho }-\widehat{Z}_{\mu \nu \rho }.$}

For nontrivial H--fields of type (\ref{ansh}), we have to re--define the
'cosmological' constant, $\lambda \rightarrow \lambda +\lambda _{\lbrack
H]}, $ respectively in the formulas (\ref{aux01})--(\ref{aux03}) and (\ref%
{tsol4}) and write, for instance, $e^{2(\lambda +\lambda _{\lbrack
H]})p^{5}} $ instead of $e^{2\lambda p^{5}}.$

\section{Conclusions}

We constructed three classes of exact solutions of Ricci flow equations in
5D spacetimes. Despite the presence in literature of rather sophisticate
mathematical considerations \cite{ricciflows} on existence of solutions
no explicit constructions attempting such constructions were performed
excepting  the linearized approach  \cite%
{crvis}, which also allows to generate exact solutions but for
lower dimensions. The Ricci flows are described by nonlinear
systems of second order parabolic differential equations which can
not approached by the usual techniques of constructing exact
solutions with diagonalizable metrics in
gravity theories. The 'anholonomic frame method' \cite%
{anhm,details1,details2} is a general geometric one allowing to construct
generic off--diagonal exact solutions which naturally can be related both to
solutions of the Einstein equations and Ricci flows.

In this paper, we analyzed flows of pp--waves on nontrivial solitonic
backgrounds of 5D spacetimes. Such equations (at least for normalized flows
and certain ansatz for string gravity) contain terms with nontrivial
cosmological constants. There is some similarity with the solutions defining
Kaigorodov spaces \cite{kaigorodov}, constructed as homogeneous Einstein
spaces induced by pp--waves propagating in anti--de Sitter space (it is
conjectured in modern literature \cite{strppwave,geg} that for a
string/M--theory on the Kaigorodov spaces a compact manifold is dual to a
conformal field theory in an infinitely--boosted frame with constant
momentum density). \footnote{%
One should be noted that nonholonomic deformations of charged Kaigorodov
spaces can be generated from the data $\underline{g}_{\alpha }$ in (\ref%
{data1}) restricted to define a particular case of exact solutions (\ref{c01})
and (\ref{c02}) defining Einstein--Maxwell configurations with nontrivial
cosmological constants as in Ref. \cite{cai}.}

The off--diagonal metric coefficients describing flows of
superpositions of pp--waves and solitons have certain analogy with
previous results on pp--waves in higher dimensions and
non--relativistic versions of Kaluza--Klein monopole \cite{duval}
locally anisotropic Taub--NUT, soliton--spinor waves, wormholes
and solitonically moving black holes in higher dimensions
\cite{vt,anhm}. Here, we emphasize that the approach elaborated in
this work contains general off--diagonal terms (so called
N--coefficients) not restricted to the Kaluza--Klein conditions
(related to linearizations and compactifications on
extra--dimension coordinates). We can consider both vacuum
gravitational pp--wave effects and Ricci flows, for instance,
self--consistently imbedded into nontrivial solitonic backgrounds
and spinor waves and/or superpositions with black hole metrics.

Our Ricci flow solutions describe nonlinear deformations and gravitational
interactions with nontrivial pp--waves and solitonic backgrounds. In a
particular case (for the ansatz for Kaigorodov solutions) they transform
into homogeneous configurations with anti--de Sitter constant, but, in
general, they are related to nonhomogeneous Einstein spaces and their string
generalizations. Such metrics are generic off--diagonal, depend on classes
of functions on one, two, three and four variables and can be with
generalized (non--Killing and non--group) symmetries.

The solitonic flows of pp--waves can be considered both on time like and/or
extra dimension coordinates, in 4D and/or 5D gravity.

\section*{Acknowledgments}

The author is grateful for
 discussions and substantial support to T. Wolf, S. Anco and M. Visinescu.
 He also thanks for support of Fields Institute.

  \end{document}